# Monte Carlo Simulation of Pulsed Laser Deposition

## Pui-Man Lam [@], S. J. Liu, and C.H. Woo

Department of Mechanical Engineering, The Hong Kong Polytechnic University, Hong Kong

## ABSTRACT

Using the Monte Carlo method, we have studied the pulsed laser deposition process at the sub-monolayer regime. In our simulations, dissociation of an atom from a cluster is incorporated. Our results indicate that the pulsed laser deposition resembles molecular beam epitaxy at very low intensity, and that it is characteristically different from molecular beam epitaxy at higher intensity. We have also obtained the island size distributions. The scaling function for the island size distribution for pulsed laser deposition is different from that of molecular beam epitaxy.

## I.INTRODUCTION

Pulsed laser deposition (PLD) is a growth technique in which the target material is ablated by a pulsed laser and then deposited in pulses on a substrate surface, i.e. many particles arrive simultaneously at the surface [1]. It is a new technique that may improve layer-by-layer growth [2, 3] and is especially suited for the growth of complex multicomponent thin films, e.g. high temperature superconductors [4], biomaterials [5], or ferroelectric films [6]. A great advantage of PLD is the conservation of the stoichiometry of virtually any target material in the deposition. Experimentally each pulse has a length of about a few nanoseconds and the time between two pulses is of the order of seconds.

Recently, Hinneman et al [7, 8] proposed a simple model for PLD. In this model the duration of a pulse is assumed to be zero and the transient enhancement of the mobility of

---

[@] On sabbatical from Department of Physics, Southern University, Baton Rouge, USA (pmlam@grant.phys.subr.edu)

freshly deposited atoms is neglected. The microstructure evolution is controlled by three parameters, namely, the intensity *I* – which is the density of particles deposited per pulse, the rate for diffusion of adatoms on the surface *D*, and the average flux of incident particles per site *F*. The atoms are deposited in pulses of intensity *I*, onto a flat substrate. The atoms diffuse on the substrate with a surface diffusion constant *D* until they meet another adatom, in which case they form a stable and immobile nucleus of a two-dimensional island, or until they attach irreversibly to the edge of an already existing island. The Ehrlich-Schwoebel barriers for atoms to descend from an island are not taken into account. Since there is no edge diffusion, the islands grow in a fractal manner before they coalesce. A similar model with a finite pulse length had been proposed previously by Jensen, Niemeyer, and Combe [9, 10]. Hinneman et al had actually restricted their PLD-simulation to a particularly simple case, namely to the limit of an infinite *D/F*, meaning that all adatoms nucleate or attach to an existing island before the next pulse arrives.

The quantity they examined was the island density $N(I,\mathbf{q})$ as a function of the intensity *I* and the coverage $\mathbf{q}$. They found that for all coverage up to the maximum coverage $\mathbf{q}_{max} = 1$ in their simulation, the island density is an increasing function of the intensity *I*. Defining the quantity $M(I,\mathbf{q}) \equiv N(I,\mathbf{q})/N(I,\mathbf{q}_{max})$ they found that the logarithm of $M(I,\mathbf{q})$ obeyed very well the scaling form $\log(M(I,\mathbf{q}))/\log(I) = g(\log(\mathbf{q})/\log(I))$ where the scaling function g(z) was very well approximated by a simple parabola $g(z) = az^2$, with *a* as a constant.

Since the irreversible model of ref. [7], resulting in fractal islands, is applicable only to very special situations, we generalize the model to include reversible processes and finite D/F. We have also obtained the island size distributions. The scaling function for the island size distribution for pulsed laser deposition is different from that of molecular beam epitaxy.

## II. KINETIC MONTE-CARLO MODEL

Here we use a more realistic model to study the PLD [7], by using the conventional kinetic Monte-Carlo approach. Atoms are deposited in regular pulses of zero duration and intensity *I*, with the average flux of incident particles per site *F*. All surface atoms, including those that are connected by nearest neighbor bonds to other atoms, can hop to nearest neighbor sites. The rate at which a surface atom with n lateral nearest neighbors can hop to a nearest neighbor site is determined by the configuration-dependent Arrhenius-type expression $k_n(T)=D\exp(-nE/k_BT)$. Here E is the potential energy of an atom with one lateral bond, $k_B$ is the Boltzmann's constant and *T* is the absolute temperature. The free-atom migration *D* is given by the expression $D = (2k_BT/h)\exp(-E_s/k_BT)$, where *h* is the Planck's constant and $E_S$ is activation energy for free adatom hopping. In our simulation we have fixed *F=0.1* (monolayer/second), and set $E_S=1.3$ eV and *E=0.3* eV, the values used by Ratsch et al. [11] in their simulation of the submonolayer island size distribution in order to compare with experimental results for Fe(100). The simulations are then performed on a square lattice of size 300x300, with results averaged over 50 runs, at various intensities for three different temperatures, *T=700K, 800K* and *850K*. The measurements always take place right before a new pulse is released. In Figures 1a and 1b we show the qualitative difference between MBE and PLD respectively in our reversible kinetic Monte-Carlo model. Both figures are for *T=800K*, flux *F=0.1* and show the typical configurations after a deposition of 0.4 monolayer. We can see that the island density is much higher for PLD even though the average flux is the same in both cases. We can also see that the islands are compact as compared to the fractal islands in the case of ref. [7].

## III. RESULTS

Figures 2a, 2b and 2c show the island density *N(I,**q**)* as a function of the coverage for temperatures *T=700K*, *800K* and *850K* and various intensities. The cases of molecular beam epitaxy (MBE) are also shown for comparison. The results for MBE are obtained when the intensity is lowered to one single atom per pulse. The values of the ratio *D/F* for the three corresponding temperatures are $10^5$, $2 \times 10^6$ and $7 \times 10^6$ respectively, in appropriate units. For all three temperatures, at least for the higher coverage, the island density increases with increasing intensity. This is plausible since for a higher intensity more atoms arrive at the surface simultaneously so that most of them can meet and form new islands before attaching to existing islands. At low intensities we expect to have MBE. But as the intensity increases there should be a crossover to PLD behavior. The crossover is expected to occur when the intensity exceeds the density $N_1$ of adatoms in MBE. We notice that there are peaks in the island density as a function of the coverage for all the intensities. As the island density increases, the islands tend to capture more and more of the diffusing adatoms, leading to a decrease of the nucleation of new islands. As the island density increases to a certain value, the capture rate of existing islands will equal to that of nucleation of new islands. After this point the island density will level off. The decrease in island density at higher values of the coverage is due to the coalescence of existing islands as their sizes increase due to adatom capture. In a point island model, the decrease of island density at higher coverage would be absent since the point islands do not coalesce. In the irreversible model of Hinneman et al [7], there are also no peaks in the island density. But it is not clear from their paper whether they had taken a point island model or not. In Figure 3 we have plotted the average single adatom density $N_1$ as a function of the coverage in MBE at the three temperatures. The peaks in the adatom density of all three curves are not higher than 0.0025 and the average adatom density over the whole coverage is about 0.001. Therefore we expect that the crossover intensity to be no higher than 0.0025 for all three temperatures and consequently for intensity higher than 0.0025, the behavior should be that of PLD, which is characterized by an increase with the island density with intensity. But since the peaks are rather narrow and the average value of $N_1$ over the whole coverage is about 0.001, we estimate the crossover intensity to be the average value of $N_1$, i.e. 0.001.

In Fig. 4a we show the peak values $N_m$ of the island density versus the intensity *I* for the three temperatures, in double logarithmic plots. We can see that $N_m$ is approximately constant below the intensity value *I* of 0.01. For *I*> 0.01, $N_m$ increases with I as $N_m \sim I^{2\nu}$, with $\nu \approx 0.125$. In Fig. 4b we show the island densities N at the fixed coverage $\theta=0.2$ versus the intensity I in double logarithmic plots. Again the behavior is similar to that of $N_m$. From figures 4a and 4b one can see that in the high intensity regime, i.e. the regime of PLD, the total island density increases as a power law of the intensity I. However, at low intensity I, the total island density is approximately constant. This is the reason why we cannot collapse the data for all intensities using a scaling function containing only one exponent. Only in the high intensity regime, where the power law holds, is the scaling good. This is different from the result of ref. [7] where only the special case D/F→∞ was studied using an irreversible model. Since for the irreversible case the crossover intensity $I_c$ for PLD goes as $I_c \sim (D/F)^{-x}$, where x>0 [7], the model studied in ref [7] is always in the PLD regime, for all intensities. It is difficult to approach the limit D/F→∞ in the reversible model, because this limit is approached in the limit of very high temperature or very low flux. In both cases, more and more time is spent in particle diffusion rather than deposition and the computation becomes increasing time consuming. However we believe that the reversible model with finite flux and large but finite D/F is more relevant to the experimental situation.

Following ref [7] we show in Figures 5a-c the double logarithmic plots of the quantity $M(I,\boldsymbol{q}) = N(I,\boldsymbol{q})/N(I,\boldsymbol{q}_{max})$ at the three temperatures, where $\boldsymbol{q}_{max}$ =0.4 is the maximum coverage in our simulation. With this definition, the rightmost points of $M(I,\boldsymbol{q})$ are collapsed to a single point. Here we find that for all three temperatures, for I ≤0.001, the data for various intensities seem to collapse into one curve. This is in agreement with the result of Figs. 4a and 4b that the island density is approximately independent of the intensity I for I ≤0.001. For all three temperatures, for I≥0.01, the data for various intensities seem to approach a different curve than the curve at low intensities. This is also in agreement with the result of Figs. 4a and 4b that the island density increases with intensity as $N \sim I^{2\nu}$, so that the ratio $M(I,\theta)$ becomes independent of I.

Again following ref [7] we show in Figures 6 a-c the quantity $-log(M(I,\boldsymbol{q}))/log(I)$ versus the quantity $-log(\boldsymbol{q})/log(I)$, for the three different temperatures. Here we can see that the curves do not scale either. This is because in ref [7] only the special case D/F→∞ was studied. Their model is therefore always in the PLD regime for all intensities. In our model with finite D/F, we are in the PLD regime only at high intensities. At low intensities we are in the MBE regime. We consider the general scaling form $-|log(M(I,\theta))|=|log(I)|^{\alpha}G(|log(\theta)|/|log(I)|^{\beta})$, with a scaling function G(x) and exponents α and β. We take the absolute values of the various quantities here because they are all less than one, which make the logarithms negative. By varying the values of the exponents α and β we can determine the best scaling function G. In Fig. 7 we show the scaling plot of $-|log(M(I,\theta))|/|log(I)|^{\alpha}$ versus $-|log(\theta)|/|log(I)|^{\beta}$, for T=800K, with α=β=1/4. The scaling appears to be rather good. One notices that the points for I < 0.001 are spread over the whole range of the figure while the points for I > 0.001 are only in the plateau of the figure. The reason for this is that for that for I > 0.001, there do not exist data points with coverage less than 0.001, because the first pulse already give a coverage of at least 0.001. Similar scaling is obtained with the same values of α and β for the other temperatures. Also, as seen from Figs. 2a-c the behavior of the island density diverges from that of MBE when the intensity is 0.001 at all three temperatures. This agrees with the average density of adatoms in the MBE growth, indicating a critical condition of crossover from MBE to PLD – that is the adatom density in MBE equals the intensity of the PLD.

In ref. [11-15] it was shown that the island size distribution N(S,θ) for the number of atoms S in an island obey the scaling $N(S,\theta) = N(S)<S>^2/\boldsymbol{q}$, where <S> is the average island size. This scaling relation can be understood as follows. Let χ(S,θ)=N(S,θ)/N(θ), where N(θ)=∫N(S,θ)dS. If χ(S,θ) obeys a scaling, then it can be written as χ(S,θ)=θ$^y$g(S/θ$^x$), where x and y are certain exponents and g(z) is the scaling function. Now let $\langle S \rangle = \theta^x$. Then one has $\chi(S,\theta) = \langle S \rangle^{y/x} g(S/\langle S \rangle)$. Putting this in the integral $\int \chi(S,\theta) dS = 1$ gives $\langle S \rangle^{1+(y/x)} \int g(z)dz = 1$. This implies $y/x = -1$ and $\int g(z)dz = 1$. We also have $\langle S \rangle = \langle S \rangle^{2+(y/x)} \int z g(z)dz$. Using $y/z = -1$, this gives $\int z g(z)dz = 1$. But $\langle S \rangle$ is also given by $\langle S \rangle = \int S N(S,\theta) dS / N(\theta) = \theta / N(\theta)$. Therefore one has the relation $N(\theta) = \theta / \langle S \rangle$. This gives $N(S,\theta) = (N(\theta)/\langle S \rangle) g(S/\langle S \rangle) = (\theta / \langle S \rangle^2) g(S/\langle S \rangle)$. In Figures 8a and 8b we show the scaled island size distribution function $N(S) \langle S \rangle^2 / q$ versus the scaled quantity $S/\langle S \rangle$ at $T=700K$, for different values of the coverage $\theta$, for the case of MBE and PLD at intensity of 0.1 respectively. To obtain these data we have averaged over 200 runs. For both cases of MBE and PLD, the data approach a scaling form [11-15], but the scaling functions are different for the two cases. The peak in the PLD distributions seems to shift towards islands of smaller size. A very similar behavior of the island size distribution, with almost the same scaling distribution, is found for PLD at intensity 0.05.

## IV. CONCLUSION

In conclusion we have simulated the PLD using a reversible kinetic Monte-Carlo model. We find that the island density increases with intensity. At very low intensity the scaling behavior is that of molecular beam epitaxy, but at higher intensity the behavior is that of PLD, characterized by increase of the island density with intensity. However, the excellent scaling form found in ref. [7] for an irreversible model, in terms of ratios of logarithms of various quantities does not seem to apply, when the reversible processes are considered. We have related the divergence to the critical condition when adatom density in MBE and intensity in PLD are equal – when the intensity of pulses is even higher, PLD behavior prevails.

## ACKNOWLEDGEMENT

The work described in this paper was supported by grants from the Research Grants Council of the Hong Kong Special Administrative Region (PolyU 1/99C, PolyU 5146/99E and PolyU 5152/00E) and US Department of Energy grant DE-FG02-97/ER25343.

5and y are certain exponents and g(z) is the scaling function. Now let $\langle S \rangle = \theta^x$. Then one has $\chi(S,\theta) = \langle S \rangle^{y/x} g(S/\langle S \rangle)$. Putting this in the integral $\int \chi(S,\theta) dS = 1$ gives $\langle S \rangle^{1+(y/x)} \int g(z)dz = 1$. This implies $y/x = -1$ and $\int g(z)dz = 1$. We also have $\langle S \rangle = \langle S \rangle^{2+(y/x)} \int z g(z)dz$. Using $y/z = -1$, this gives $\int z g(z)dz = 1$. But $\langle S \rangle$ is also given by $\langle S \rangle = \int S N(S,\theta) dS / N(\theta) = \theta / N(\theta)$. Therefore one has the relation $N(\theta) = \theta / \langle S \rangle$. This gives $N(S,\theta) = (N(\theta)/\langle S \rangle) g(S/\langle S \rangle) = (\theta / \langle S \rangle^2) g(S/\langle S \rangle)$. In Figures 8a and 8b we show the scaled island size distribution function $N(S) \langle S \rangle^2 / q$ versus the scaled quantity $S/\langle S \rangle$ at $T=700K$, for different values of the coverage $\theta$, for the case of MBE and PLD at intensity of 0.1 respectively. To obtain these data we have averaged over 200 runs. For both cases of MBE and PLD, the data approach a scaling form [11-15], but the scaling functions are different for the two cases. The peak in the PLD distributions seems to shift towards islands of smaller size. A very similar behavior of the island size distribution, with almost the same scaling distribution, is found for PLD at intensity 0.05.

## IV. CONCLUSION

In conclusion we have simulated the PLD using a reversible kinetic Monte-Carlo model. We find that the island density increases with intensity. At very low intensity the scaling behavior is that of molecular beam epitaxy, but at higher intensity the behavior is that of PLD, characterized by increase of the island density with intensity. However, the excellent scaling form found in ref. [7] for an irreversible model, in terms of ratios of logarithms of various quantities does not seem to apply, when the reversible processes are considered. We have related the divergence to the critical condition when adatom density in MBE and intensity in PLD are equal – when the intensity of pulses is even higher, PLD behavior prevails.

## ACKNOWLEDGEMENT

The work described in this paper was supported by grants from the Research Grants Council of the Hong Kong Special Administrative Region (PolyU 1/99C, PolyU 5146/99E and PolyU 5152/00E) and US Department of Energy grant DE-FG02-97/ER25343.

## REFERENCES

1. D. B. Chrisey and G. K. Hubler, Pulsed Laser Deposition of Thin Films (John Wiley & Son, New York, 1994).
2. H. Jenniches, M. Klaua, H. Höchle, J. Kirschner, Appl. Phys. Lett. **69**, 3339 (1996).
3. H. Huang, G. H. Gilmer, T. Diaz de la Rubia, J. Appl. Phys. **84**, 3636 (1998).
4. J. T. Cheung, P. E. D. Morgan, D. H. Lowndes and X.-Y. Zheng, Appl. Phys. Lett. **62**, 2045 (1993).
5. C. M. Cotell, D. B. Chrisey, K. S. Grabowski and J. S. Sprague, J. Appl. Biomat. **3**, 87 (1992).
6. R. Ramesh, A. Inam, W. K. Chan and B. Wilkens, Science **252**, 944 (1991).
7. B. Hinneman, H. Hinrichsen and D. E. Wolf, Phys. Rev. Lett. **87**, 135701(2001)
8. B. Hinnemann, F. Westerhoff and D. E. Wolf, Cond-mat/0104329, 18 April 2001.

FIGURE CAPTIONS.

Fig. 1a. Typical island configuration for MBE after deposition of 0.4 monolayer, at T=800K, F=0.1

Fig. 1b. Typical island configuration for PLD after deposition of 0.4 monolayer, at T=800K, F=0.1 and intensity I=0.1

Fig. 2a: Island density versus coverage for T=700K

Fig. 2b: Island density versus coverage for T=800K

Fig. 2c: Island density versus coverage for T=850K

Fig. 3: Single adatom density as function of coverage for the three temperatures

Fig. 4a: Peak values $N_m$ of the island density versus intensity I for the three temperatures

Fig. 4b: Island density N at coverage $\theta=0.2$ versus intensity I for the three temperatures

Fig. 5a: Island density for T=700K rescaled so that all curves terminate at rightmost point

Fig. 5b: Same as Fig. 5a, but for T=800K

Fig. 5c: Same as Fig. 5a, but for T=850K

Fig. 6a: Quantity $-\log(M(I))/\log(I)$ versus quantity $-\log(\theta)/\log(I)$, for T=700K

Fig. 6b: Same as Fig. 6a, but for T=800K

Fig. 5c: Same as Fig. 6a, but for T=850K

Fig. 7: Scaling plot of $|\log(M(I,\theta)|/|\log(I)|^\alpha$ versus $-|\log(\theta)|/\log(I)|^\beta$, for T=800K, with $\alpha=\beta=1/4$.

Fig. 8a: Scaling plot of the island size distribution for MBE

Fig. 8b: Scaling plot of the island size distribution for PLD

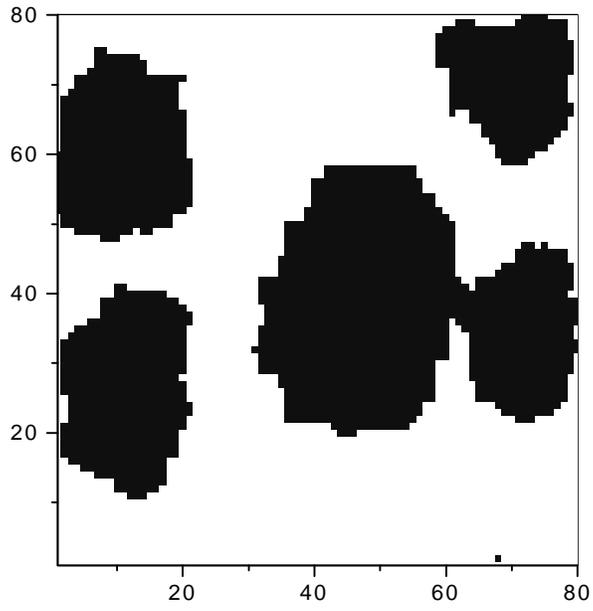

Fig. 1a

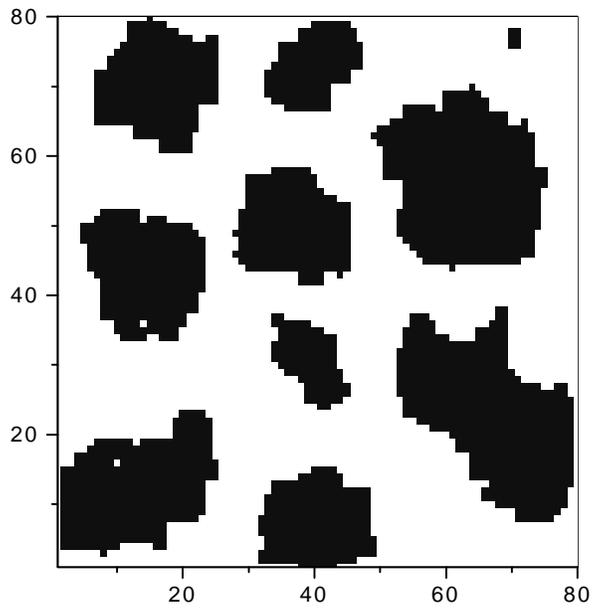

Fig. 1b

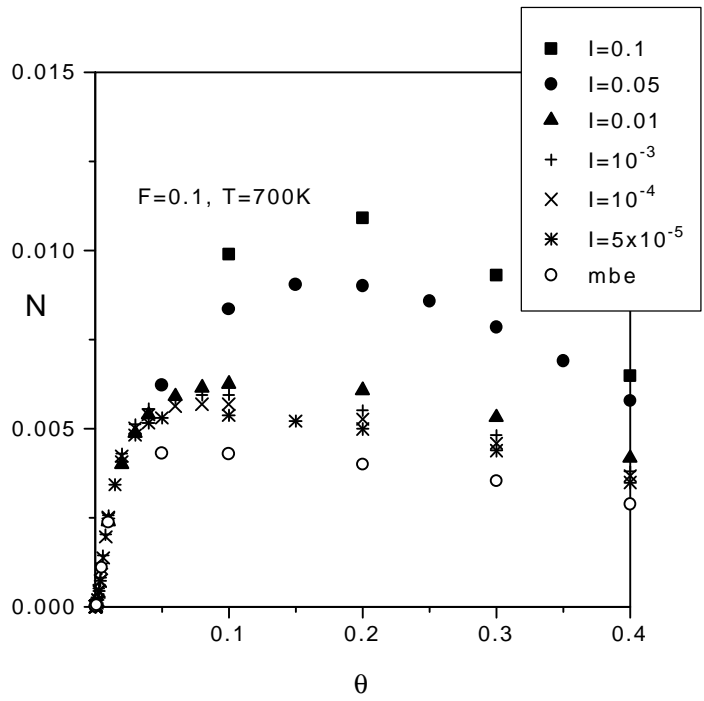

Fig. 2a

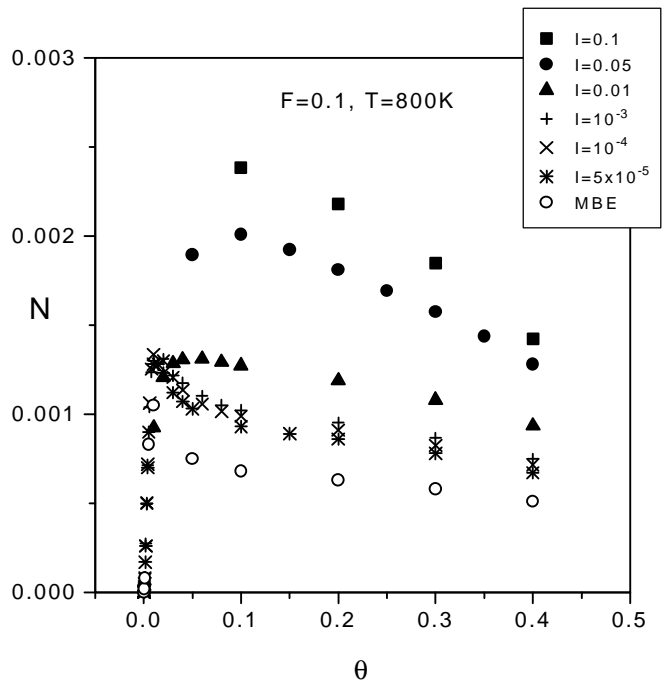

Fig. 2b

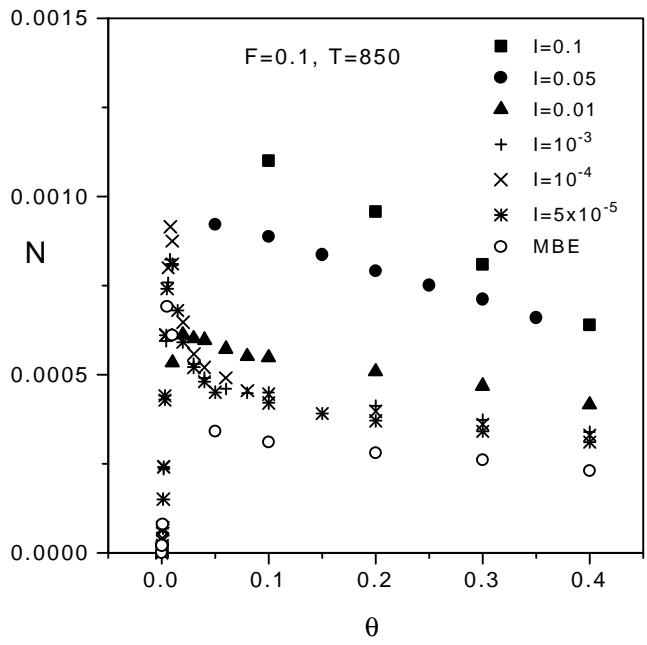

Fig. 2c

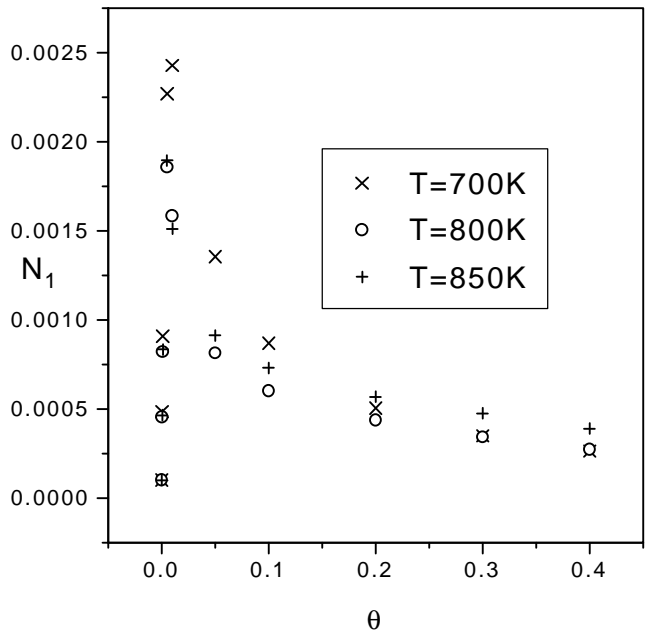

Fig. 3

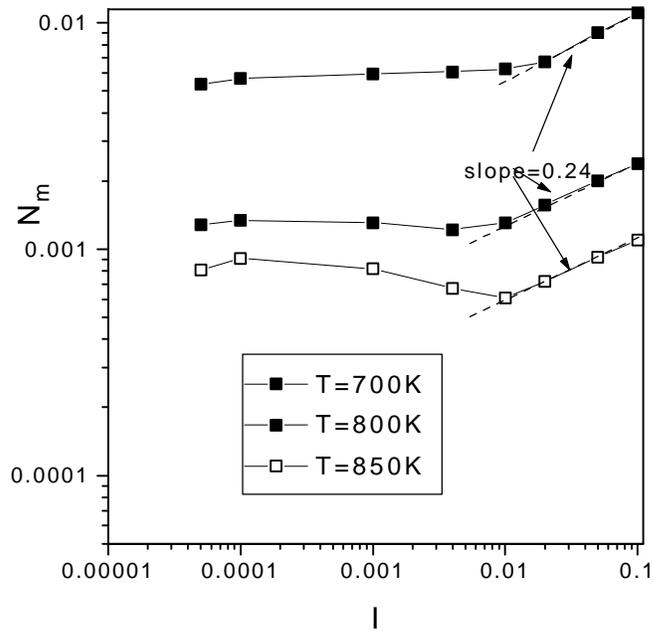

Fig. 4a

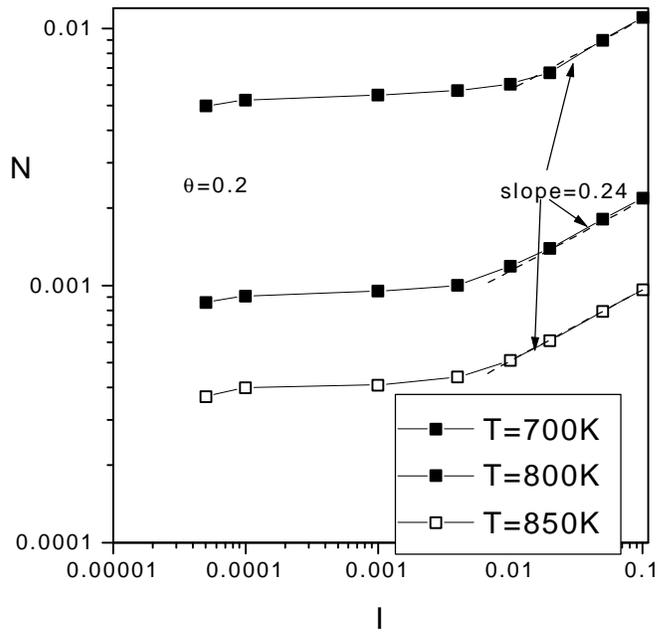

Fig. 4b

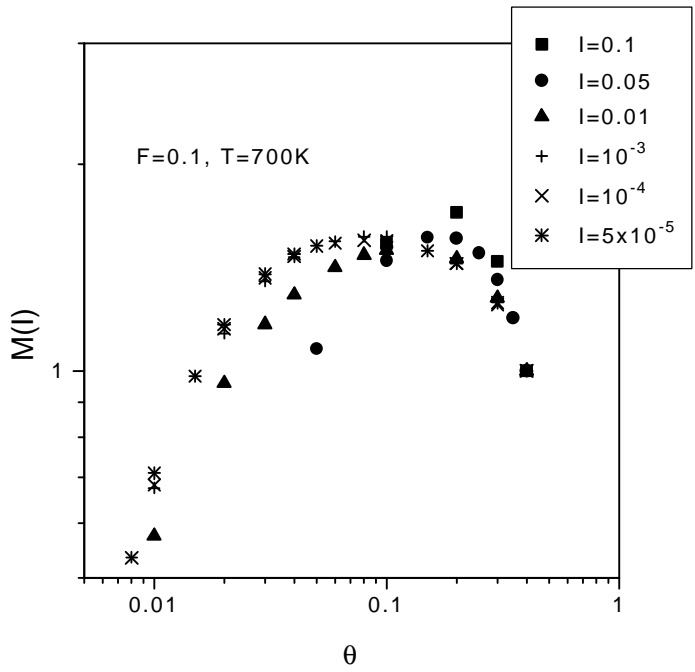

Fig. 5a

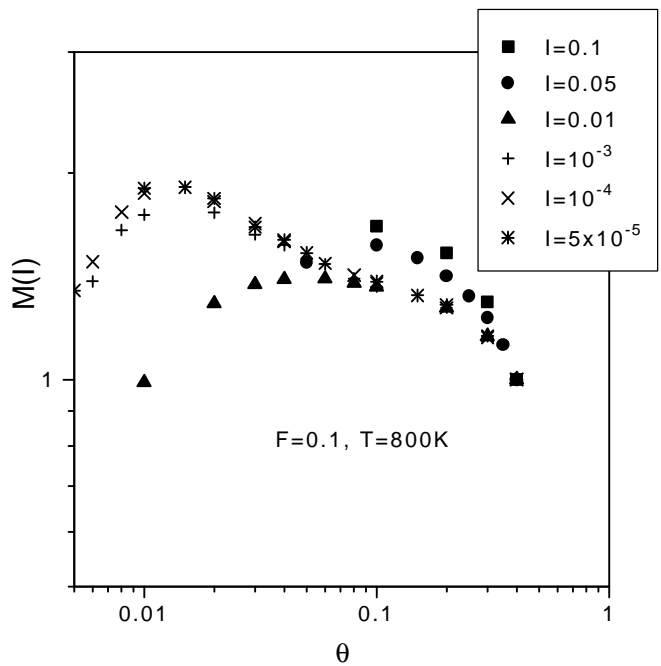

Fig. 5b

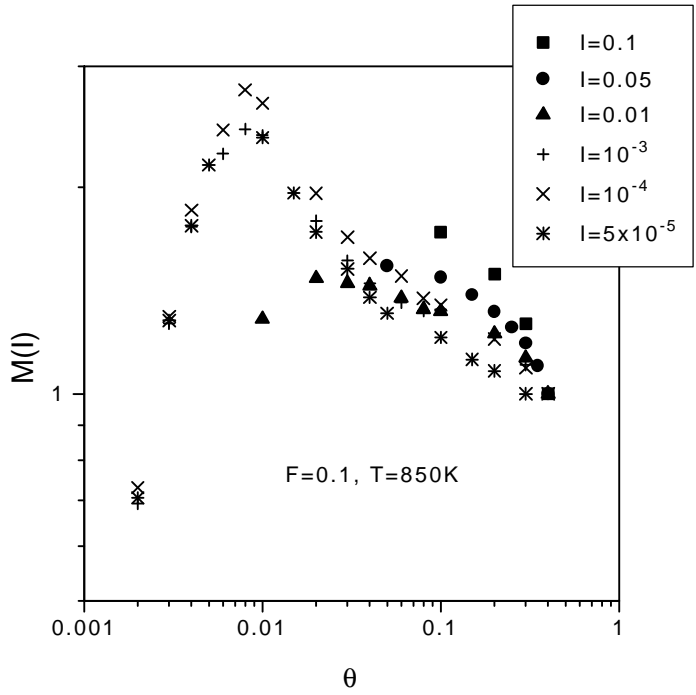

Fig. 5c

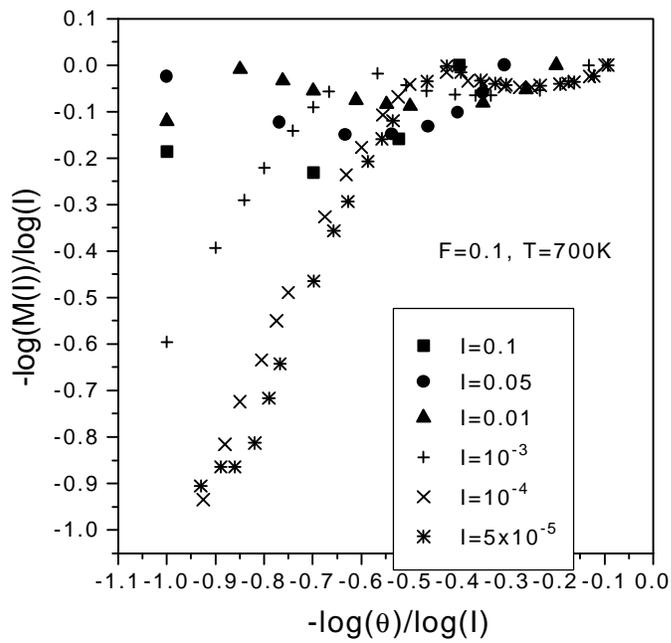

Fig. 6a

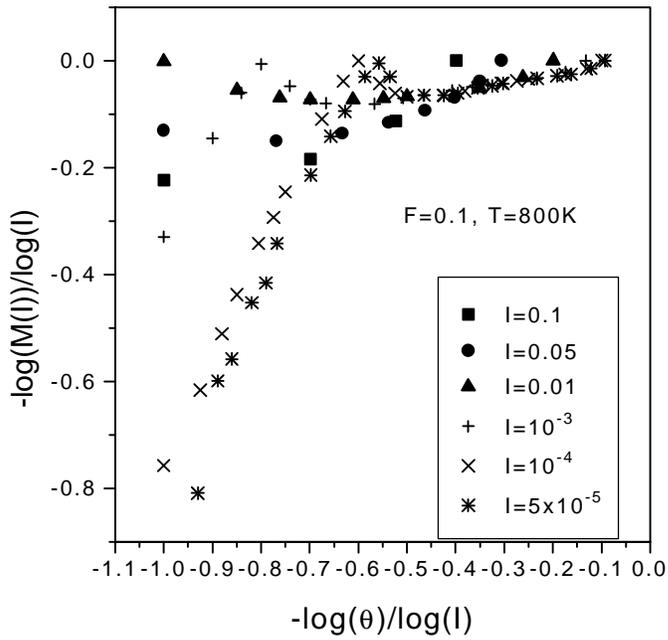

Fig. 6b

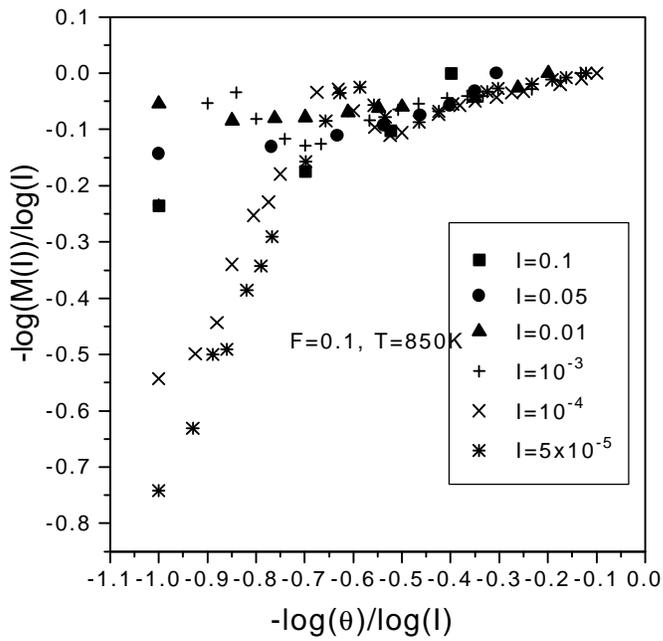

Fig. 6c

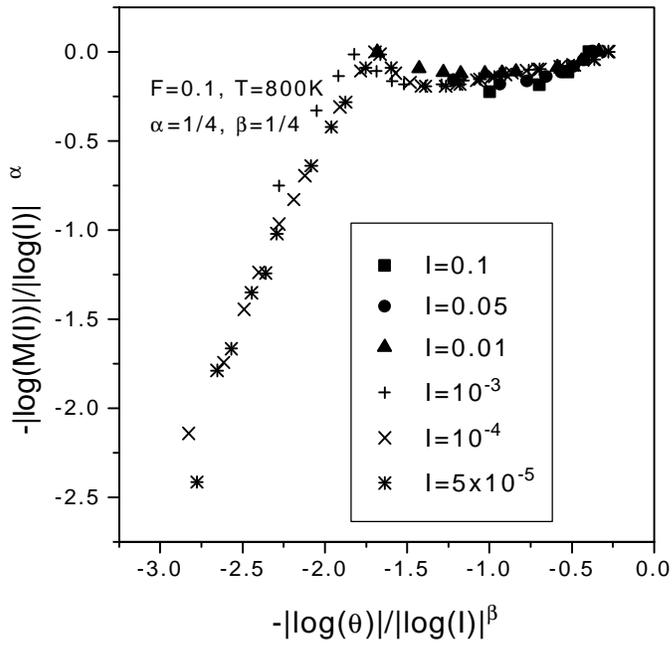

Fig. 7

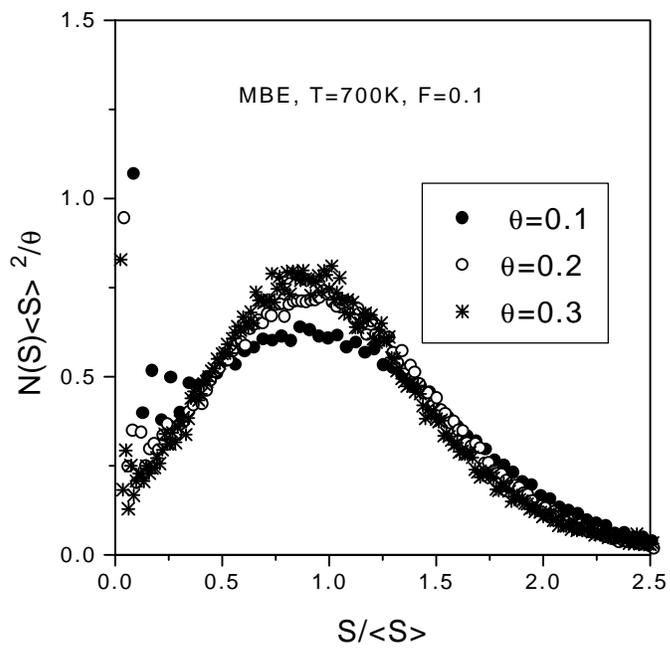

Fig. 8a

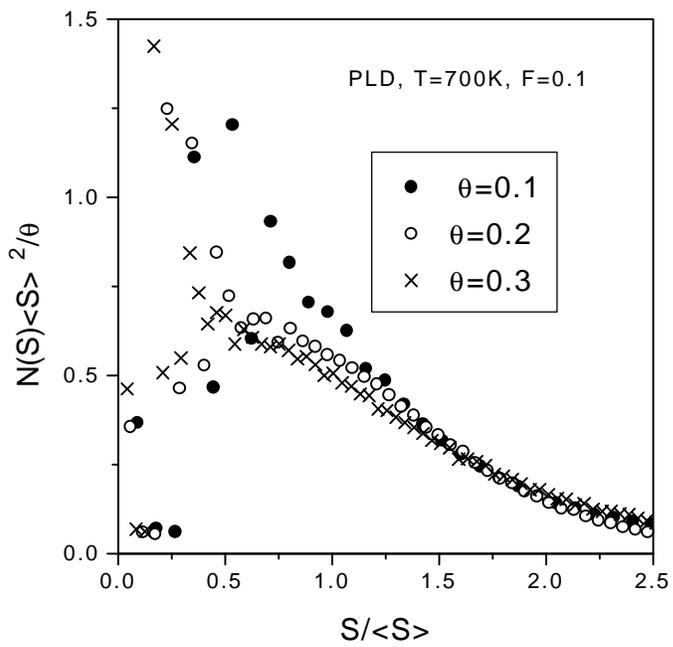

Fig. 8b